\documentclass[a4paper,12pt]{article}

\usepackage{epsfig}
\usepackage{amssymb}
\usepackage{amsfonts}
\usepackage{amsmath}
\usepackage{slashed}
\usepackage{euscript}
\usepackage{verbatim}
\usepackage{latexsym}
\usepackage{graphicx}
\usepackage{caption}
\usepackage{float}
\usepackage{subcaption}
\usepackage{wrapfig}
	\usepackage[T1]{fontenc}
	\usepackage{tikz}
	\usetikzlibrary{decorations.pathmorphing}
\usepackage{tikz}
\usetikzlibrary{shapes.geometric, arrows,patterns,snakes}
\tikzstyle{ellip} = [ellipse, minimum width=3cm, minimum height=1cm,text centered, draw=black]

\newskip\humongous \humongous=0pt plus 1000pt minus 1000pt

\newif\ifdtup

\catcode`\@=11

\@addtoreset{equation}{section}

\def\@normalsize{\@setsize\normalsize{15pt}\xiipt\@xiipt
\abovedisplayskip 14pt plus3pt minus3pt%
\belowdisplayskip \abovedisplayskip
\abovedisplayshortskip \z@ plus3pt%
\belowdisplayshortskip 7pt plus3.5pt minus0pt}

\def\small{\@setsize\small{13.6pt}\xipt\@xipt
\abovedisplayskip 13pt plus3pt minus3pt%
\belowdisplayskip \abovedisplayskip
\abovedisplayshortskip \z@ plus3pt%
\belowdisplayshortskip 7pt plus3.5pt minus0pt
\def\@listi{\parsep 4.5pt plus 2pt minus 1pt
     \itemsep \parsep
     \topsep 9pt plus 3pt minus 3pt}}

\relax

\catcode`@=12

\catcode`\@=11

\def\section{\@startsection{section}{1}{\z@}{3.5ex plus 1ex minus
   .2ex}{2.3ex plus .2ex}{\large\bf}}

\def\SymBoxes#1#2#3#4{\newdimen\un@t \un@t#3%
\raisebox{#1}{\rule{#2\un@t}{#4}\hskip-#2\un@t
\@tempdimb\un@t \advance\@tempdimb by-#4\@tempcntb#2\relax%
\@whilenum{\@tempcntb>0}\do{
\rule{#4}{\un@t}\hskip\@tempdimb \advance\@tempcntb by\m@ne}%
\hskip-#2\un@t \rule[\un@t]{#2\un@t}{#4}%
\rule[\un@t]{#4}{#4}\hskip-#4
\rule{#4}{\un@t}}\hskip-#4}                

\begin{document}

\newcommand{\beq}{\begin{equation}}
\newcommand{\eeq}{\end{equation}}
\newcommand{\bea}{\begin{eqnarray}}
\newcommand{\eea}{\end{eqnarray}}
\newcommand{\beas}{\begin{eqnarray*}}
\newcommand{\eeas}{\end{eqnarray*}}
\newcommand{\defi}{\stackrel{\rm def}{=}}
\newcommand{\non}{\nonumber}
\newcommand{\bquo}{\begin{quote}}
\newcommand{\enqu}{\end{quote}}
\renewcommand{\(}{\begin{equation}}
\renewcommand{\)}{\end{equation}}
\def \eqn#1#2{\begin{equation}#2\label{#1}\end{equation}}
\def\IZ{{\mathbb Z}}
\def\IR{{\mathbb R}}
\def\IC{{\mathbb C}}
\def\IQ{{\mathbb Q}}
\def\de{\partial}
\def\Tr{ \hbox{\rm Tr}}
\def\H{ \hbox{\rm H}}
\def\HE{ \hbox{$\rm H^{even}$}}
\def\HO{ \hbox{$\rm H^{odd}$}}
\def\K{ \hbox{\rm K}}
\def\Im{ \hbox{\rm Im}}
\def\Ker{ \hbox{\rm Ker}}
\def\const{\hbox {\rm const.}}
\def\o{\over}
\def\im{\hbox{\rm Im}}
\def\re{\hbox{\rm Re}}
\def\bra{\langle}\def\ket{\rangle}
\def\Arg{\hbox {\rm Arg}}
\def\Re{\hbox {\rm Re}}
\def\Im{\hbox {\rm Im}}
\def\exo{\hbox {\rm exp}}
\def\diag{\hbox{\rm diag}}
\def\longvert{{\rule[-2mm]{0.1mm}{7mm}}\,}
\def\a{\alpha}
\def\dag{{}^{\dagger}}
\def\tq{{\widetilde q}}
\def\p{{}^{\prime}}
\def\W{W}
\def\N{{\cal N}}
\def\hsp{,\hspace{.7cm}}

\def\br{\nonumber\\}
\def\IZ{{\mathbb Z}}
\def\IR{{\mathbb R}}
\def\IC{{\mathbb C}}
\def\IQ{{\mathbb Q}}
\def\IP{{\mathbb P}}
\def \eqn#1#2{\begin{equation}#2\label{#1}\end{equation}}

\newcommand{\sgm}[1]{\sigma_{#1}}
\newcommand{\idd}{\mathbf{1}}

\newcommand{\C}{\ensuremath{\mathbb C}}
\newcommand{\Z}{\ensuremath{\mathbb Z}}
\newcommand{\R}{\ensuremath{\mathbb R}}
\newcommand{\rp}{\ensuremath{\mathbb {RP}}}
\newcommand{\cp}{\ensuremath{\mathbb {CP}}}
\newcommand{\vac}{\ensuremath{|0\rangle}}
\newcommand{\vact}{\ensuremath{|00\rangle}                    }
\newcommand{\oc}{\ensuremath{\overline{c}}}
\begin{titlepage}
\begin{flushright}
CHEP XXXXX
\end{flushright}
\bigskip
\def\thefootnote{\fnsymbol{footnote}}

\begin{center}
{\Large
{\bf Quantum Chaos and Holographic Tensor Models
}
}
\end{center}

\bigskip
\begin{center}
{\large  Chethan KRISHNAN$^a$\footnote{\texttt{chethan.krishnan@gmail.com}}, Sambuddha SANYAL$^b$\footnote{sambuddha.sanyal@icts.res.in}, \vspace{0.15in} \\ P. N. Bala SUBRAMANIAN$^a$\footnote{\texttt{pnbalasubramanian@gmail.com}}}
\vspace{0.1in}

\end{center}

\renewcommand{\thefootnote}{\arabic{footnote}}

\begin{center}
$^a$ {Center for High Energy Physics,\\
Indian Institute of Science, Bangalore 560012, India}

$^b$ {International Center for Theoretical Sciences,\\
Tata Institute of Fundamental Research, Bangalore 560089, India}\\

\end{center}

\noindent
\begin{center} {\bf Abstract} \end{center}


A class of tensor models were recently outlined as potentially calculable examples of holography: their perturbative large-$N$ behavior is similar to the Sachdev-Ye-Kitaev (SYK) model, but they are fully quantum mechanical (in the sense that there is no quenched disorder averaging). These facts make them intriguing tentative models for quantum black holes. In this note, we explicitly diagonalize the simplest non-trivial Gurau-Witten tensor model and study its spectral and late-time properties. We find parallels to (a single sample of) SYK where some of these features were recently attributed to random matrix behavior and quantum chaos. In particular, after a running time average, the spectral form factor exhibits striking qualitative similarities to SYK. But we also observe that even though the spectrum has a unique ground state, it has a huge (quasi-?)degeneracy of intermediate energy states, not seen in SYK. If one ignores the delta function due to the degeneracies however, there is level repulsion in the unfolded spacing distribution hinting chaos. Furthermore, the spectrum has gaps and is not (linearly) rigid. The system also has a spectral mirror symmetry which we trace back to the presence of a unitary operator with which the Hamiltonian anticommutes. We use it to argue that to the extent that the model exhibits random matrix behavior, it is controlled not by the Dyson ensembles, but by the BDI (chiral orthogonal) class in the Altland-Zirnbauer classification.







\vspace{1.6 cm}
\vfill

\end{titlepage}

\setcounter{footnote}{0}

\section{Motivation and Conclusions}
\noindent

In the holiday wish-list \cite{Polchinski} of a devout Holographer, one might very well find a theory that exhibits (a) solvability in the large-$N$ limit, (b) maximal chaos \cite{bound}, and (c) emergent conformal symmetry in the infrared. A theory with these properties would be a potential candidate for a controllable holographic model for quantum black holes. At first glance, these demands together might seem forbiddingly constraining\footnote{In particular, Nature probably does not owe us solvability.}, but a remarkable theory that passes all three criteria is known: this is the 0+1 dimensional model of Sachdev, Ye and Kitaev (SYK) \cite{SYK}. See also \cite{old, coredump}.

The SYK model has ``quenched disorder'', which means that it is a theory whose correlation functions are to be considered {\em after}\footnote{Of course, one can also consider the theory where the couplings realize only a single element of the ensemble. Indeed, we will see that this could actually be interesting for our discussions, see also section 8 of \cite{crowdsourced}. But the exact solvability at large-$N$ of SYK is unfortunately and crucially tied to the ensemble average.} an average over an ensemble of couplings. This means that the SYK (ensemble-averaged) correlation functions cannot themselves be interpreted as those of a true quantum system, and therefore one might worry about the lessons one can extract about the quantum behavior of black holes by studying them. 

As an antidote to this, Witten proposed \cite{Witten} a class of tensor models  (building on the work of Gurau and collaborators \cite{guraudump}) which have the same large-$N$ ``melonic'' behavior  \cite{Gurau} as the SYK model and therefore shares its nice features, but does not require a quench. We will call these models and their relatives \cite{klebanov} Holographic Tensor Models (HTM). In this paper, we will explicitly solve the simplest\footnote{The effective $N$ for this model turns out to be 32, which makes it comparable to the $N=32$ version of the SYK model that already exhibits \cite{crowdsourced} many large-$N$ features.} non-trivial Gurau-Witten tensor model.

Our interest in this problem is directly motivated by the work of \cite{crowdsourced, garcia}, who studied spectral properties of the SYK model and showed that it exhibits various features that are characteristic of random matrices and quantum chaos \cite{Haake, Shukla}. In particular, \cite{crowdsourced} considered a specific function constructed from the spectrum of the theory\footnote{They call this function the Spectral Form Factor (SFF), and we will adopt this terminology. See \cite{sffpapers} for various previous discussions on the SFF.} and showed that a specific dip-ramp-plateau structure in its time-dependence is a signature also shared by random matrices in the appropriate ensembles. This statement is true without further qualifications for the SYK model after the ensemble average. But even for a single realization of SYK, this statement holds 
after a running time average\footnote{See section 8 of \cite{crowdsourced} and our discussions later for a precise definition of the running time average.}.  In this paper, we will show that striking qualitative similarities with this picture exist also in the Gurau-Witten tensor model (after the running time average to kill the late-time fluctuations). This is interesting because unlike in the (single realization of the) SYK model, the coupling here is a single (dimensionful) number, not ${\cal O}(N^4)$ numbers each chosen from a Gaussian distribution. This result is indicative that despite this, there is randomness and chaos in the system.

We will also see however that there are some interesting differences between the tensor model and SYK. One of the most striking features is that the tensor model has what looks \footnote{To within our numerical error, which we cannot entirely rule out: We have precision up to $\approx 10$  decimals.} like a huge degeneracy in the middle of the energy spectrum, as well as moderate degeneracy elsewhere. The ground state however, is unique. It is tempting to speculate that such a large degeneracy has to do with the entropy of black hole states in the theory  \cite{QFTBH}, and that it has something to say about the zero modes of the broken emergent reparametrization \cite{Maldacena} in the IR. But we emphasize that the true ground state is unique and {\em not} degenerate. See \cite{ads2} for discussions on the emergent reparametrization in the holographically dual AdS$_2$. Since the system is fermionic, it is plausible that the ``half-filled" state should be viewed as the Fermi surface and states above and below it are to be thought of as particles and holes. This is especially likely in light of the fact that the spectrum has a particle-hole-like mirror symmetry as we will discuss. 

It will be very interesting to understand these degeneracies in terms of some underlying symmetry of the Gurau-Witten model. Note that for some values of $N$ the SYK model also had degeneracies because of fermionic symmetries related to Bott periodicity. But these were degeneracies that affected every level. Here on the other hand, the degeneracies affect every state except the most positive and most negative energy states, but the actual degeneracies are different for each level. The maximum degeneracy occurs at $E=0$. But we also note that extremely finely spaced quasi-degeneracies near zero are known in some condensed matter systems \cite{Sambuddha}. Also, in related uncolored tensor models of \cite{klebanov}, at least in some cases we have checked \cite{Vamsi} that many of the same features we see here remain, but the degeneracy is lifted. See the footnote in our final section for some more comments on this.

Interestingly, once we remove the degeneracies and look at the (unfolded) level spacing distribution $P(s)$, we find distinct evidence that the system shows level repulsion at low $s$ indicative of chaotic dynamics. Another feature  we see is that the spectrum has gaps in it, especially close to the mid-levels of the energy. In this, and the fact that the spectrum has no (linear) spectral rigidity, the holographic tensor model is distinct from  SYK \cite{crowdsourced}. Spectral rigidity (see eg. \cite{Berry}) is a measure of how much the integrated density of states (sometimes called the spectral staircase) deviates from a linear fit: it seems from our figure 4 that our spectrum does not have linear spectral rigidity\footnote{We have checked that the spectrum in a single sample of SYK can be fit quite well with a linear fit.}. The lack of spectral rigidity is responsible for the difference between the ``ramp" parts of our plots of the SFF, see figures 5 and 6, and those in \cite{crowdsourced}. In the holographic tensor model, we find that the plot rises up quite quickly after the dip to a plateau, in other words the ramp (to the extent that it is well-defined) is quite steep. This is perhaps not surprising because in \cite{crowdsourced} it was shown that their slow ramp structure is related to rigidity of the spectrum. We emphasize however that the statistics we have for the eigenvalues is relatively small, and that these claims should be taken with a pinch of salt. We note that the late-time plateau is also related to the level repulsion that wee see in the spacing distribution \cite{crowdsourced}. Quantum chaos, random matrix-like aspects and eigenstate thermalization in certain gapped systems has been studied in \cite{Rigol}.

Yet another striking feature of the spectrum is that it has a mirror symmetry, by which we mean that the energy levels come in pairs around the center as
\bea
(E_0+E_n, \ \ E_0-E_n).
\eea
The midpoint energy is $E_0=0$ and it is at that energy that we see the huge degeneracy. The presence of spectral mirror symmetry is an indication that the system has a discrete symmetry which we will discuss in detail later. We will see that it can be traced to the existence of a unitary operator that {\em anti-}commutes with the Hamiltonian \cite{Witten}. We will explicitly construct this operator for our Gurau-Witten model.  Together with the presence of a Particle-Hole Symmetry operator which has already been identified for SYK and SYK-like models like ours \cite{Ludwig, Fu, crowdsourced}, this helps us fix the symmetry class of the theory. We will find that the symmetry class is the so-called BDI class in the 10-fold classification of Altland and Zirnbauer \cite{AZ}. This means that unlike the SYK models which were controlled (depending on the parity properties of $N$) by the Gaussian Unitary, Orthogonal and Symplectic ensembles of Dyson, the random matrix behavior of this model is likely to be controlled by the chiral Gaussian Orthogonal Ensemble. We leave a detailed study of these and numerous other interesting questions for future work, some of which we comment on in a final section.

\section{The Holographic Tensor Model}

The general Gurau-Witten tensor model contains $q=D+1$ real fermionic fields 
\bea
\psi_{a, i_{a0}...\slashed{i_{aa}}...i_{aD}}
\eea
where $a, b \in \{0,1,...,D\}$ are called colors, and each of the $i_{am}$'s run from $1,..,n$, where $n$ is independent of $D$. The notation $\slashed{i_{aa}}$ means that $i_{aa}$ is omitted in the indices. The transformation property of the index $i_{am}$ is what defines the symmetry group of the theory, and it is fixed as follows. First we define a group $G_{ab}=O(n)$ for each unordered pair $(a,b)$ of distinct elements in $\{0,1,...,D\}$. This means that upto an overall discrete group that we will not keep track of in this paper, the symmetry group of the theory is
\bea
G \sim O(n)^{D(D+1)/2}
\eea
Now the index $i_{am}$ is thought of as transforming in the vector represnetation of $G_{am}$ for each $m \neq a$. Since there are $D$ groups $G_{ab}$ with $a \neq b$ for a given $a$, each $\psi_a$ has $n^D$ components. Now the Gurau-Witten action is written as
\bea
S_{GW}=\int dt \Bigl(\frac{i}{2}\psi_i\partial_{t}\psi_i -  \frac{i^{(D+1)/2}J}{n^{D(D-1)/4}} \ \psi_{0} \psi_{1} .. \psi_{D} \Bigr) \label{GWD}
\eea
where we have suppressed the contractions in the interaction term. Since $a$ runs from 0 to $D$, the total number of real fermions in the theory is $N=(D+1)n^D$. This is the $N$ that is relevant for large $N$, in the sense of comparison to SYK: remember the $q$ in SYK is $(D+1)$ here. The sum over $i$ in the kinetic term is from 1 to $N$. It should be clear that because the index structure of each $\psi_a$ is explicitly constructed to reflect the rest of the fields in the theory, the contraction structure when explicitly written out is a bit of a mess; see eg. \cite{Gurau} for the explicit form of the action. We will only discuss the simplest Gurau-Witten theories where it will be straightforward to write down the contractions by inspection. We also note that the scaling in the coupling $J$ is introduced so that we have well-defined large-$N$ limit. We will often set this $J$ to unity, taking advantage of the fact that it is dimension-ful.

Lets start with the simplest theories, where $D=1$. In this case, we have two sets of fields: $\psi_0$ transforming as a vector under $G_{01}=O(n)$ and $\psi_1$ transforming as a vector under $G_{10}=G_{01}$. This means that the theory is an $O(n)$ theory and explicitly we have
\bea
S_{GW}^{D=1}=\int dt \Bigl(\frac{i}{2}\psi_a^i\partial_{t}\psi_a^i -  {i J} \ \psi_{0}^i\psi_1^i \Bigr) \label{GW1}
\eea
where all indices are explicit and repeated indices are summed over their appropriate ranges. This theory is trivially solved for any value of $n$ because it is free after an appropriate diagonalization in field space: we will not present the details. Essentially identical discussions can be found in eg. \cite{Maldacena, crowdsourced} in the context of SYK.

Since the Lagrangian has to be a boson, the next simplest example corresponds to $D=3$.  
Some index chasing and being careful about the locations of contractions shows that the explicit action is given by:
\bea
S_{GW}^{D=3}= \int dt \Bigl(\frac{i}{2}\psi_{a}^{ijk}\partial_{t}\psi_{a}^{ijk} + \frac{J}{n^{3/2}} \psi_{0}^{ijk} \psi_{1}^{ilm} \psi_{2}^{njm} \psi_{3}^{nlk} \Bigr) \label{GW3}
\eea
The specific contraction structure that we follow here follows a similar contraction in \cite{klebanov}. But one can check (and we have, explicitly) that other consistent contractions also lead to identical eigenvalue spectra: this is expected because this just affects the ordering of the assignment of gamma matrices to the spinors (see next section). 
The theory has an $O(n)^6$ symmetry group, and the number of fermions in the theory is $4 n^3$. The case $n=2$ will be the subject matter of most of our discussions. 

\section{The $D=3, n=2$ Gurau-Witten Hamiltonian} 

Our goal in this paper is to diagonalize the Hamiltonian corresponding to \eqref{GW3} and use it to investigate whether the system exhibits any features of chaos/random-matrix behavior. 

The canonical anti-commutation relations of the theory immediately lead to the Clifford algebra
\bea
\{\psi_a^{ijk}, \psi_b^{lmn} \} = \delta_{ab}\delta^{il}\delta^{jm}\delta^{kn}.
\eea
This means that we can realize the fermion operators in 0+1 dimensions as Euclidean Gamma matrices\footnote{The nomenclature here in the condensed matter literature is a bit confusing to the high energy theorist. To emphasize the obvious: there are no genuine spinors in 0+1 D. What is meant by a fermion in 0+1 dimensional quantum mechanics is an operator that satisfies the Clifford algebra, in other words a gamma matrix. The dimensionality of the Clifford representation is a choice one has the freedom to make, independent of the spacetime dimension which is of course 0+1. In the SYK model for instance, this choice of $N$ gets interpreted as the number of lattice sites.} of $SO(N)=SO((D+1) n^D)$. The dimension of the spinors on which they act grow exponentially fast in $N$, so if we want to have any chance of solving these on a computer, we need to stick to low values for $D$ and $n$: the upper limit for $N$ that is tractable on a computer is about 32, 34, ... from what we see in papers on the subject. Quite fortunately, we find that the first non-trivial value for $N$ in the Gurau-Witten model corresponds to $n=2$ which yields $N=32$. This is the model we will solve in this paper. 

Note that we got lucky: the next lowest GW model is computationally inaccessible and requires too much RAM to store the matrices (at least by our resources and skills in computing), as we will discuss later. It is also fortuitous that the solvable $N$ is not too low! If it were, we could not legitimately hope to reasonably claim that we are seeing hints of any large-$N$ physics. As it happens, $N=32$ happens to fall in the right range, and it also happens to be around the upper boundary of $N$ considered in the work of \cite{crowdsourced}.

\subsection{Friendly and Really-Real Spinor Representations}

The gamma matrices we will need are those of $SO(32)$ which means they are going to be 65536 $\times$ 65536 matrices. To solve them with our computing resources, we found it best to work not with the standard representation of gamma matrices which are complex, but instead with a real symmetric representation. The fact that such a representation exists is guaranteed in $N=0 \mod 8$ dimensions. We will use the so-called friendly representation of gamma matrices \cite{sugra} where the gamma matrices are "really real" in $N=0 \mod 8$ dimensions. To construct them systematically, we adopt the following recipe. We first construct Euclidean gamma matrices $E_i$ in $N=8$ 
\bea
E_{1} &=& \sgm{1} \otimes \idd \otimes \idd\otimes \idd, \nonumber\\
E_{2} &=& \sgm{3} \otimes \idd \otimes \idd\otimes \idd, \nonumber\\
E_{3} &=& \sgm{2} \otimes \sgm{2} \otimes \sgm{1}\otimes \idd, \nonumber\\
E_{4} &=& \sgm{2} \otimes \sgm{2} \otimes \sgm{3}\otimes \idd, \nonumber\\
E_{5} &=& \sgm{2} \otimes \sgm{1} \otimes \idd \otimes \sgm{2}, \nonumber\\
E_{6} &=& \sgm{2} \otimes \sgm{3} \otimes \idd \otimes \sgm{2}, \nonumber\\
E_{7} &=& \sgm{2} \otimes \idd \otimes \sgm{2} \otimes \sgm{1}, \nonumber\\
E_{8} &=& \sgm{2} \otimes \idd \otimes \sgm{2} \otimes \sgm{3}.
\eea
These can be explicitly checked to satisfy the Clifford algebra. Together with the definition
\bea
E_{*} &=& E_{1}\dots E_{8} = \sgm{2} \otimes \sgm{2} \otimes \sgm{2} \otimes \sgm{2},
\eea 
now we can follow the recipe \cite{sugra} 
\bea
\gamma^{\mu} &=& \tilde{\gamma}^{\mu} \otimes E_{*}, \ \ \mu = 0,1,\dots, D-1,\nonumber \\ 
\gamma^{D-1+i} &=& \idd \otimes E_{i} ,\ \ \quad i=1,2\dots,8.
\eea
to construct gamma matrices in $D+8$ dimensions starting from those in $D$. Starting from eight dimensions and doing this three times we get from $N=$8 to 16 to 24 to 32, which is the case we want. These gamma matrices are real and symmetric. 

\subsection{Hamiltonian in Terms of Gamma Matrices}

\begin{figure}[h]
\centering
\includegraphics[width=0.4\textwidth]{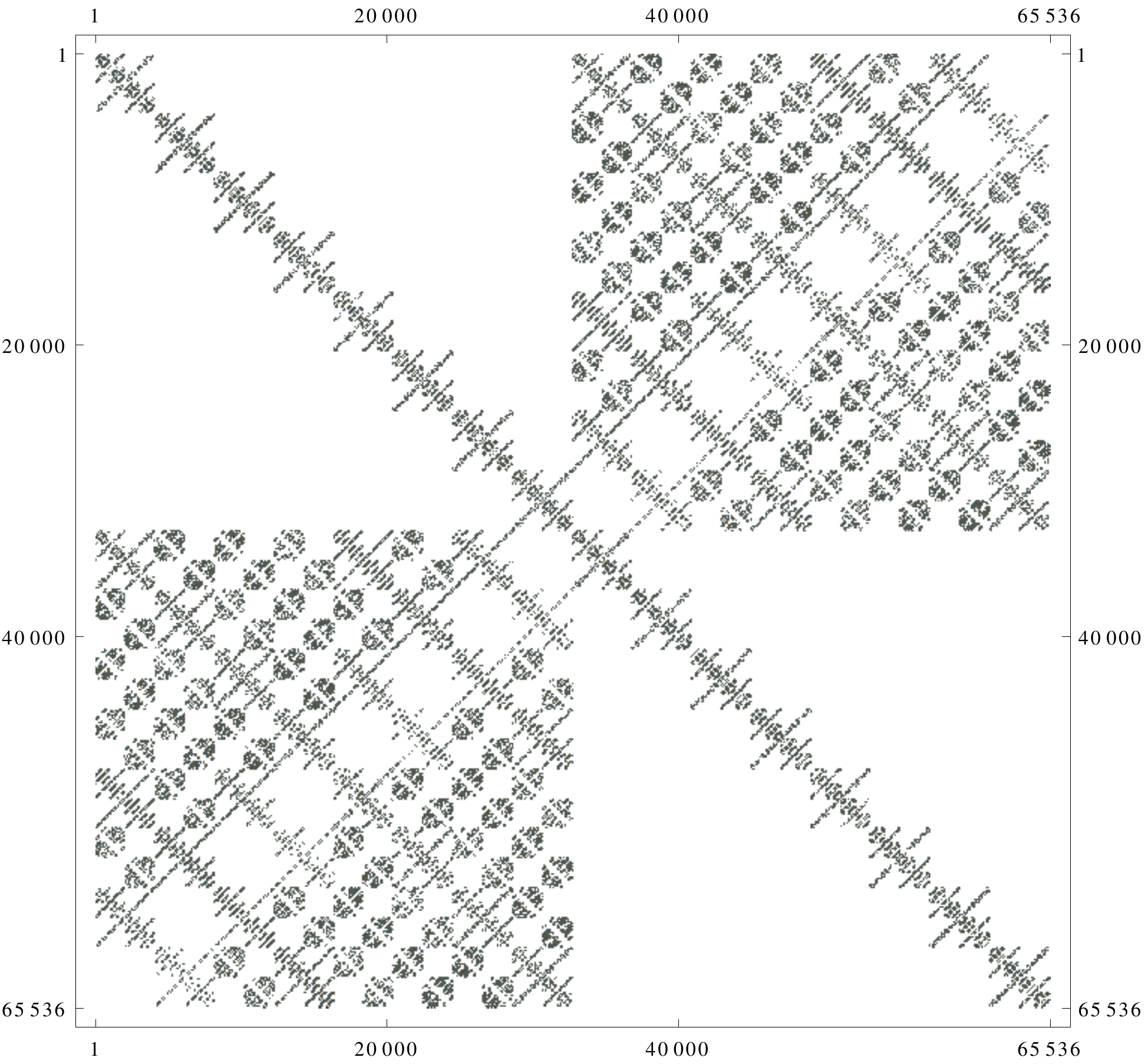}
\caption{The MatrixPlot of Hamiltonian \eqref{Ham}.}
\label{hamplot}
\end{figure}

\begin{figure}[h]
\centering
\includegraphics[width=0.4\textwidth]{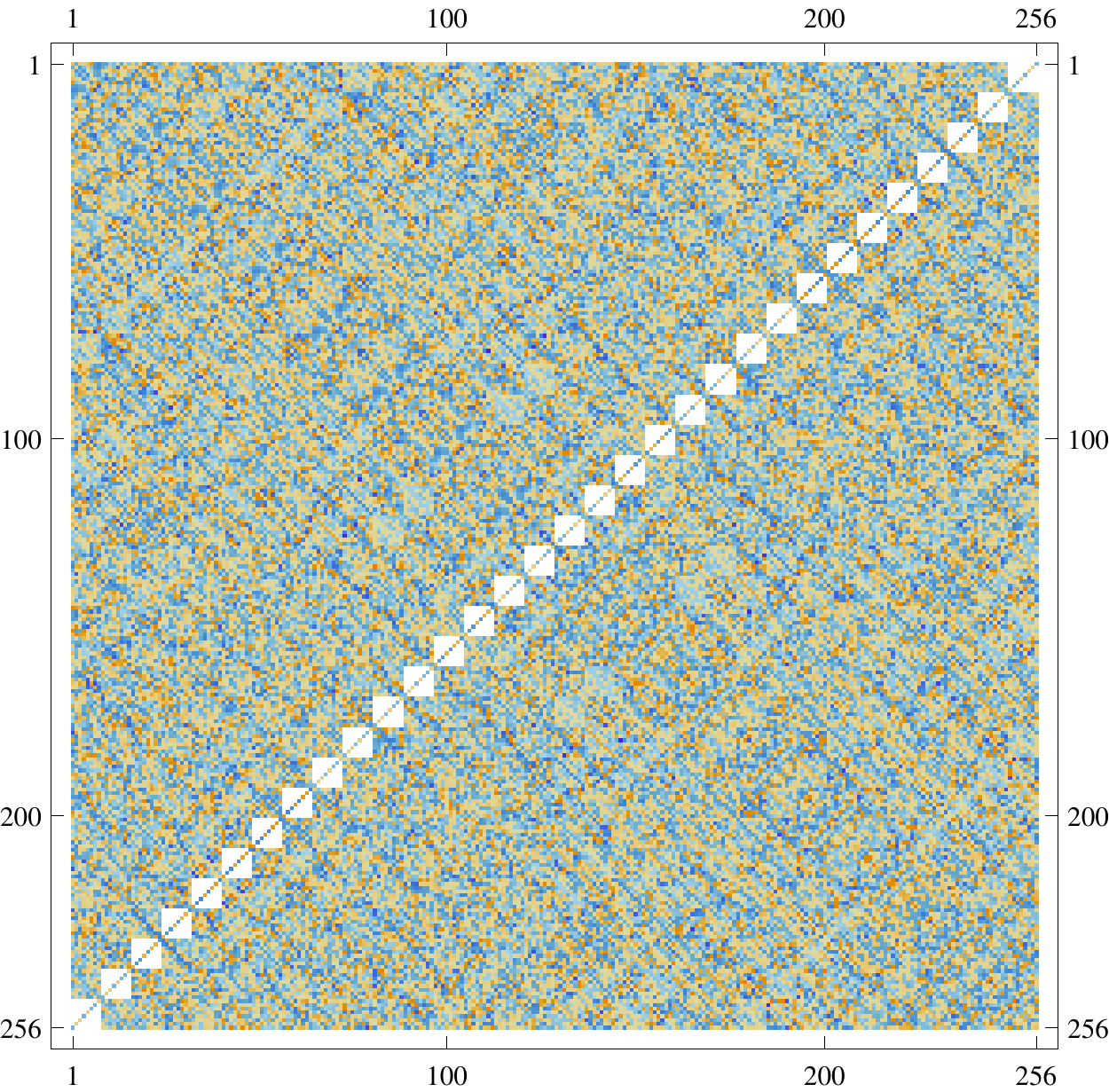}
\caption{The MatrixPlot of an SYK Hamiltonian for a single draw from the ensemble. We are considering the case $N=16$, with really real Gamma matrices.}
\label{hamplotsyk}
\end{figure}

Using these gamma matrices as our definition of the fermions, we can explicitly write out the Gurau-Witten Hamiltonian in terms of the $SO(32)$ gamma matrices. The result is a bit cumbersome:
\bea
H &=& \frac{J}{\sqrt{8}}\Bigl(\gamma_{1}\gamma_{9}\gamma_{17}\gamma_{25}+\gamma_{1}\gamma_{9}\gamma_{21}\gamma_{29}+\gamma_{1}\gamma_{10}\gamma_{18}\gamma_{25}+\gamma_{1}\gamma_{10}\gamma_{22}\gamma_{29}+\gamma_{1}\gamma_{11}\gamma_{17}\gamma_{27} \nonumber \\
&&+\gamma_{1}\gamma_{11}\gamma_{21}\gamma_{31}+\gamma_{1}\gamma_{12}\gamma_{18}\gamma_{27}+\gamma_{1}\gamma_{12}\gamma_{22}\gamma_{31}+\gamma_{2}\gamma_{9}\gamma_{17}\gamma_{26}+\gamma_{2}\gamma_{9}\gamma_{21}\gamma_{30}\nonumber \\
&&+\gamma_{2}\gamma_{10}\gamma_{18}\gamma_{26}+\gamma_{2}\gamma_{10}\gamma_{22}\gamma_{30}+\gamma_{2}\gamma_{11}\gamma_{17}\gamma_{28}+\gamma_{2}\gamma_{11}\gamma_{21}\gamma_{32}+\gamma_{2}\gamma_{12}\gamma_{18}\gamma_{28}\nonumber \\
&&+\gamma_{2}\gamma_{12}\gamma_{22}\gamma_{32}+\gamma_{3}\gamma_{9}\gamma_{19}\gamma_{25}+\gamma_{3}\gamma_{9}\gamma_{23}\gamma_{29}+\gamma_{3}\gamma_{10}\gamma_{20}\gamma_{25}+\gamma_{3}\gamma_{10}\gamma_{24}\gamma_{29}\nonumber \\
&&+\gamma_{3}\gamma_{11}\gamma_{19}\gamma_{27}+\gamma_{3}\gamma_{11}\gamma_{23}\gamma_{31}+\gamma_{3}\gamma_{12}\gamma_{20}\gamma_{27}+\gamma_{3}\gamma_{12}\gamma_{24}\gamma_{31}+\gamma_{4}\gamma_{9}\gamma_{19}\gamma_{26}\nonumber \\
&&+\gamma_{4}\gamma_{9}\gamma_{23}\gamma_{30}+\gamma_{4}\gamma_{10}\gamma_{20}\gamma_{26}+\gamma_{4}\gamma_{10}\gamma_{24}\gamma_{30}+\gamma_{4}\gamma_{11}\gamma_{19}\gamma_{28}+\gamma_{4}\gamma_{11}\gamma_{23}\gamma_{32}\nonumber \\
&&+\gamma_{4}\gamma_{12}\gamma_{20}\gamma_{28}+\gamma_{4}\gamma_{12}\gamma_{24}\gamma_{32}+\gamma_{5}\gamma_{13}\gamma_{17}\gamma_{25}+\gamma_{5}\gamma_{13}\gamma_{21}\gamma_{29}+\gamma_{5}\gamma_{14}\gamma_{18}\gamma_{25}\nonumber \\
&&+\gamma_{5}\gamma_{14}\gamma_{22}\gamma_{29}+\gamma_{5}\gamma_{15}\gamma_{17}\gamma_{27}+\gamma_{5}\gamma_{15}\gamma_{21}\gamma_{31}+\gamma_{5}\gamma_{16}\gamma_{18}\gamma_{27}+\gamma_{5}\gamma_{16}\gamma_{22}\gamma_{31}\nonumber \\
&&+\gamma_{6}\gamma_{13}\gamma_{17}\gamma_{26}+\gamma_{6}\gamma_{13}\gamma_{21}\gamma_{30}+\gamma_{6}\gamma_{14}\gamma_{18}\gamma_{26}+\gamma_{6}\gamma_{14}\gamma_{22}\gamma_{30}+\gamma_{6}\gamma_{15}\gamma_{17}\gamma_{28}\nonumber \\
&&+\gamma_{6}\gamma_{15}\gamma_{21}\gamma_{32}+\gamma_{6}\gamma_{16}\gamma_{18}\gamma_{28}+\gamma_{6}\gamma_{16}\gamma_{22}\gamma_{32}+\gamma_{7}\gamma_{13}\gamma_{19}\gamma_{25}+\gamma_{7}\gamma_{13}\gamma_{23}\gamma_{29}\nonumber \\
&&+\gamma_{7}\gamma_{14}\gamma_{20}\gamma_{25}+\gamma_{7}\gamma_{14}\gamma_{24}\gamma_{29}+\gamma_{7}\gamma_{15}\gamma_{19}\gamma_{27}+\gamma_{7}\gamma_{15}\gamma_{23}\gamma_{31}+\gamma_{7}\gamma_{16}\gamma_{20}\gamma_{27}\nonumber \\
&&+\gamma_{7}\gamma_{16}\gamma_{24}\gamma_{31}+\gamma_{8}\gamma_{13}\gamma_{19}\gamma_{26}+\gamma_{8}\gamma_{13}\gamma_{23}\gamma_{30}+\gamma_{8}\gamma_{14}\gamma_{20}\gamma_{26}+\gamma_{8}\gamma_{14}\gamma_{24}\gamma_{30}\nonumber \\
&&+\gamma_{8}\gamma_{15}\gamma_{19}\gamma_{28}+\gamma_{8}\gamma_{15}\gamma_{23}\gamma_{32}+\gamma_{8}\gamma_{16}\gamma_{20}\gamma_{28}+\gamma_{8}\gamma_{16}\gamma_{24}\gamma_{32}\Bigr) \label{Ham}
\eea
This object is what we will diagonalize and study in the upcoming sections. All its elements are either +1, -1 or zero. The matrix is largely sparse, and it is useful for some of our purposes later to have an idea about the distribution of its non-trivial matrix elements, so we plot it in Figure \ref{hamplot}. It is evident that it has some interesting (almost fractal-like) structure. It is also interesting to note that the result of a single draw of the SYK ensemble  (with the same really real Gamma matrices) results in a Hamiltonian which looks a lot more ``random" and less sparse in appearance. We present its sparseness structure in Figure \ref{hamplotsyk} for comparison. It is worth noting that the non-zero elements of such an SYK Hamiltonian are randomly distributed numbers, whereas the elements of the GW Hamiltonian are  +1, -1 or zero. And yet, we will see that it produces features of randomness. This is not unfamiliar in the case of condensed matter systems where eigenvalue spectra of adjacency matrices can give rise to randomness.

We have diagonalized the Hamiltonian above numerically, and we report on various aspects of the result in the next section.

\section{Numerical Results}

We first present the spectrum, and then in the subsequent subsections present qualitative comparisons to various spectral properties of the SYK model as well as to hints of random matrix-like behavior and chaos. We also mention the differences  from SYK.

\subsection{The Eigenvalue Spectrum}

\begin{figure}[h]
\centering
\includegraphics[width=0.5\textwidth]{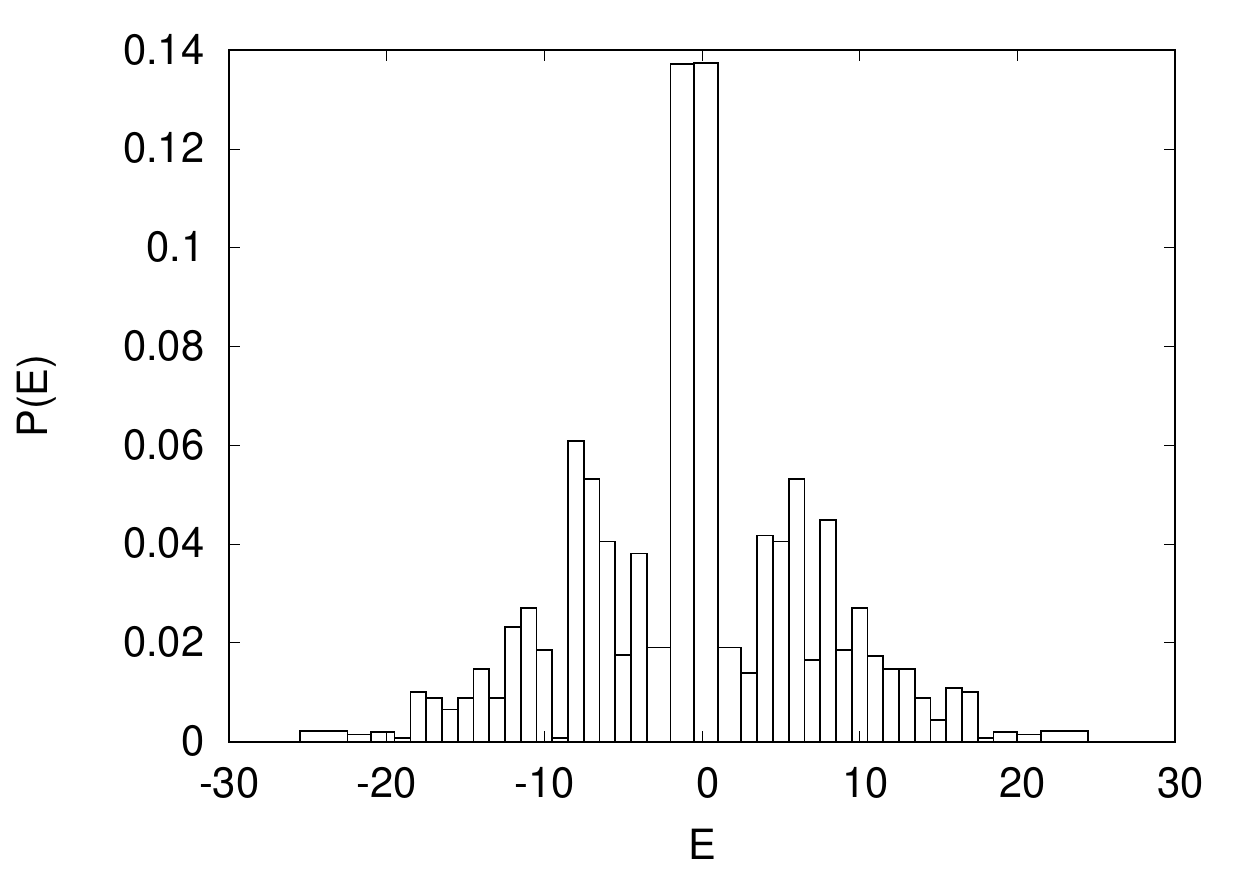}
\caption{The density of states. The d.o.s is symmetric: the slight asymmetry is an artifact of the binning of the eigenvalues.}
\label{dosplot}
\end{figure}

The density of states is plotted in Figure \ref{dosplot}.  It has a multi-peak structure that differs from the SYK single draw case \cite{Maldacena}.  
We also note that the spectrum is {\em exactly} symmetric around $E=0$. We will have more to say about this in the next section, but for now, we note that an {\em approximate} symmetry of this type existed also in (a single draw of) the SYK spectrum as well: see figure 13 in \cite{Maldacena}. We also note that the ground state is unique and has no degeneracies, but there is a huge degeneracy around $E=0$ (within our numerical precision).
\begin{figure}[h]
\centering
\includegraphics[width=0.5\textwidth]{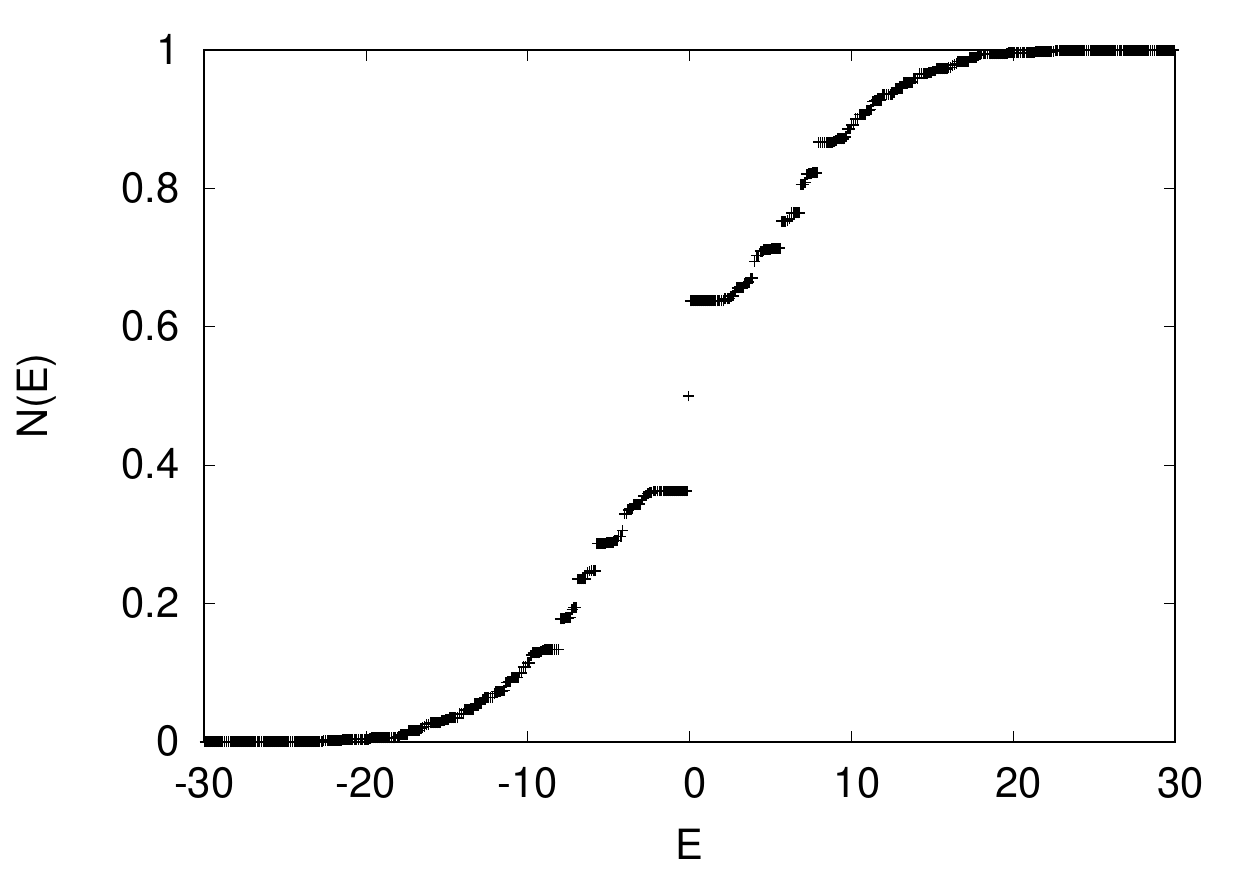}
\caption{The integrated density of states. The jump around zero is a result of the degeneracy at $E=0$.}
\label{idosfullplot}
\end{figure}

\subsection{Spectral Form Factor}

The plots of the spectral form factor, which is defined \cite{crowdsourced} as 
\bea
F_\beta(t)=\frac{|Z(\beta,t)|^2}{|Z(\beta)|^2}
\eea
with 
\bea
Z(\beta,t)\equiv {\rm Tr}\bigl(e^{-(\beta+it)H}\bigr)
\eea
was used as a measure of the random-matrix-like behavior of the SYK model. A dip-ramp-plateau structure in the theory was argued to be evidence for this. The work of \cite{crowdsourced} mostly focused on the ensemble-averaged case, but it was also noted that a running time average in the single draw case results in qualitatively similar features. 
\begin{figure}[h]
\centering
\includegraphics[width=0.5\textwidth]{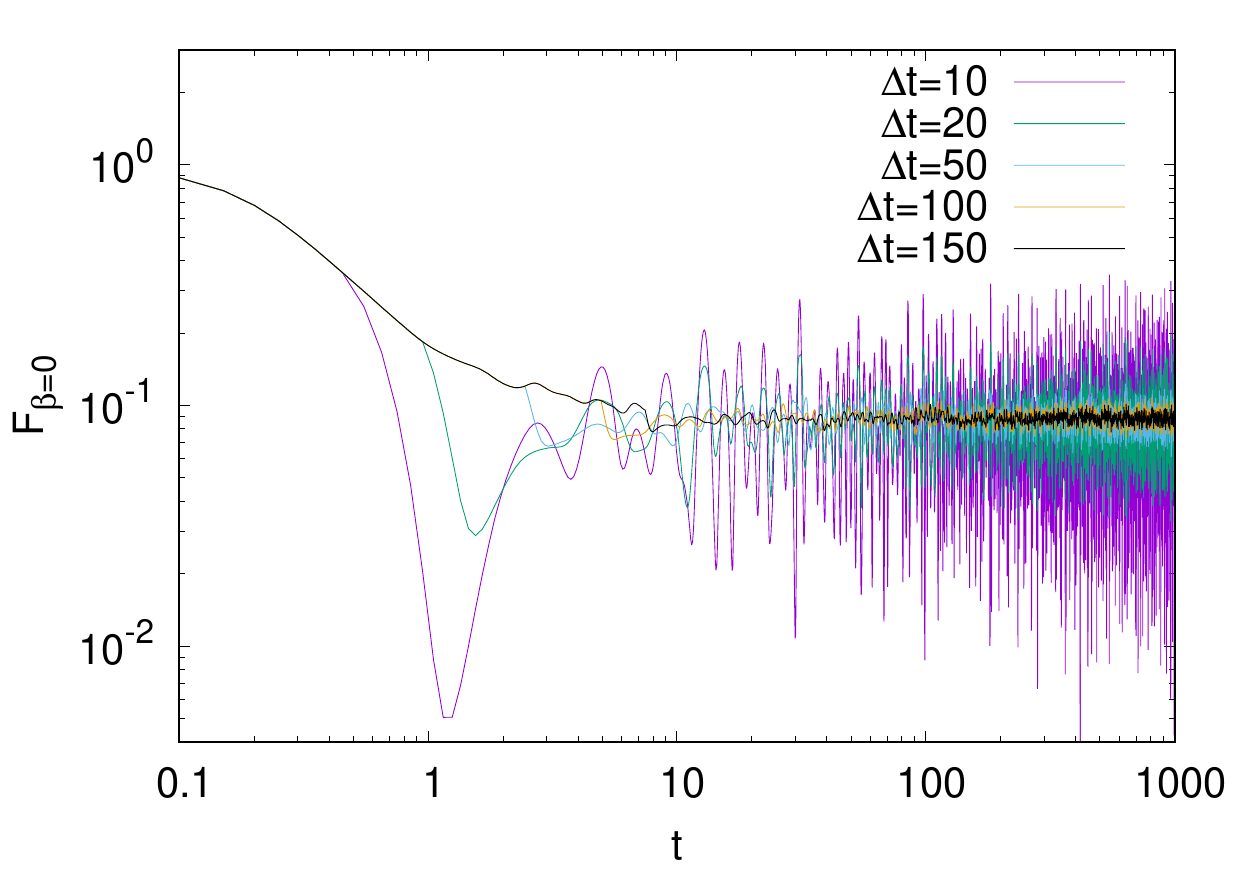}
\caption{The SFF for $\beta=0$}
\label{F0plot}
\end{figure}

\begin{figure}[h]
\centering
\includegraphics[width=0.5\textwidth]{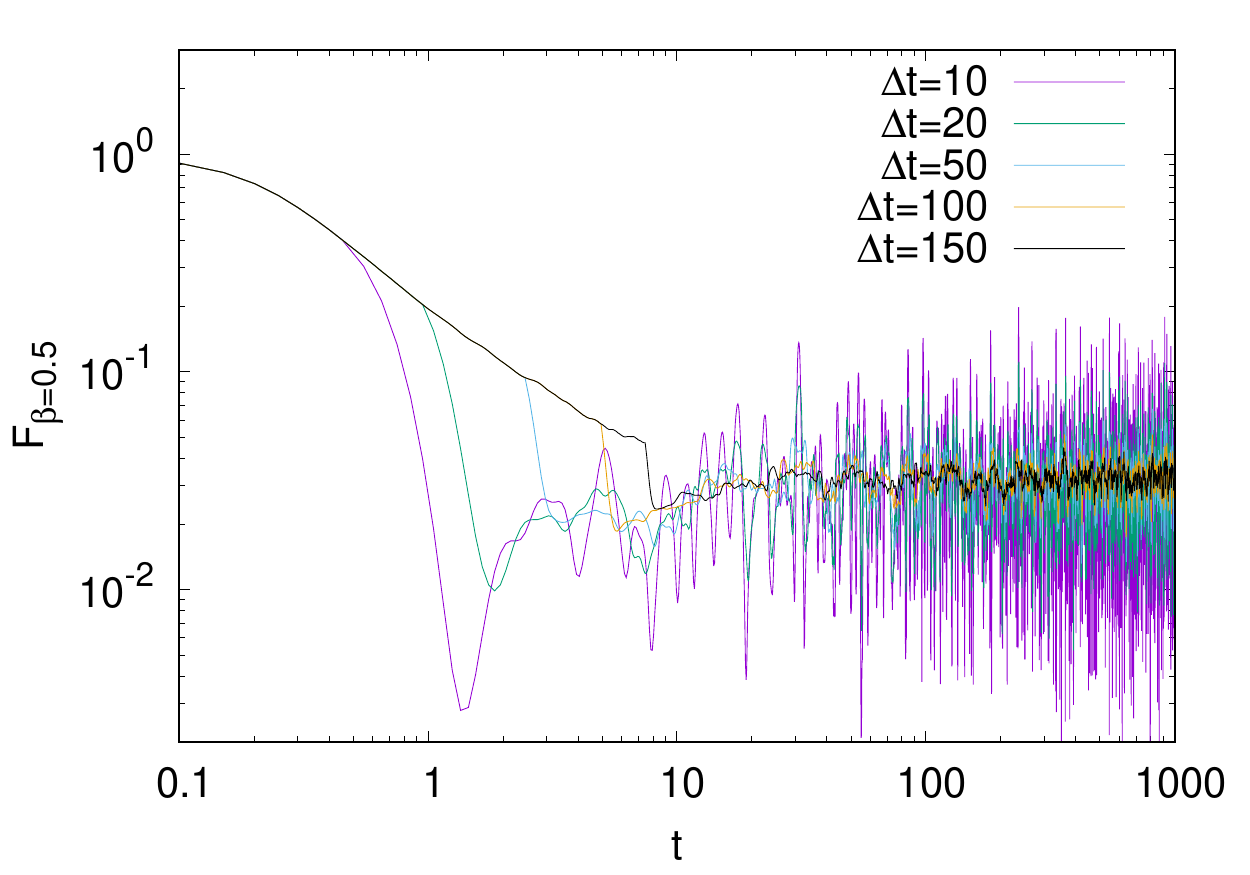}
\caption{The SFF for $\beta=0.5$}
\label{F2plot}
\end{figure}
We have computed the same quantity in the Gurau-Witten theory and we report the plots after a running time average. This means we plot a sliding window average with fixed time windows given by $\Delta t$. The averaging times $\Delta t$  are quoted in the figures. We see a pattern that is quite parallel to that found in \cite{crowdsourced}. Note also that our ramp is steeper than the one found there. We also note (as observed in \cite{Craps}) that there is some tension between increasing the averaging window and the existence of the ramp. 

\subsection{Level Repulsion} 

Once the degeneracies are removed (so that the delta function at the origin of the level spacing distribution goes away), we find that the level spacing distribution $P(s)$ shows distinct signs of level repulsion. 

To see this, we first have to unfold the spectrum (see \cite{garcia} and refernces therein). In integrable systems, the unfolded level spacing distribution typically shows a Poisson distribution steadily increasing as $s \rightarrow 0$. The absence of this, and a turnaround in the distribution close to zero is called level repulsion and is often taken as an indicator of chaotic behavior in the dynamics. In the plot \ref{unfold}, we see distinct evidence for this type of level repulsion\footnote{We have truncated the plot at large $s$ to avoid featuring edge effects: these are not relevant for seeing the level repulsion.}. 
\begin{figure}[h]
\centering
\includegraphics[width=0.5\textwidth]{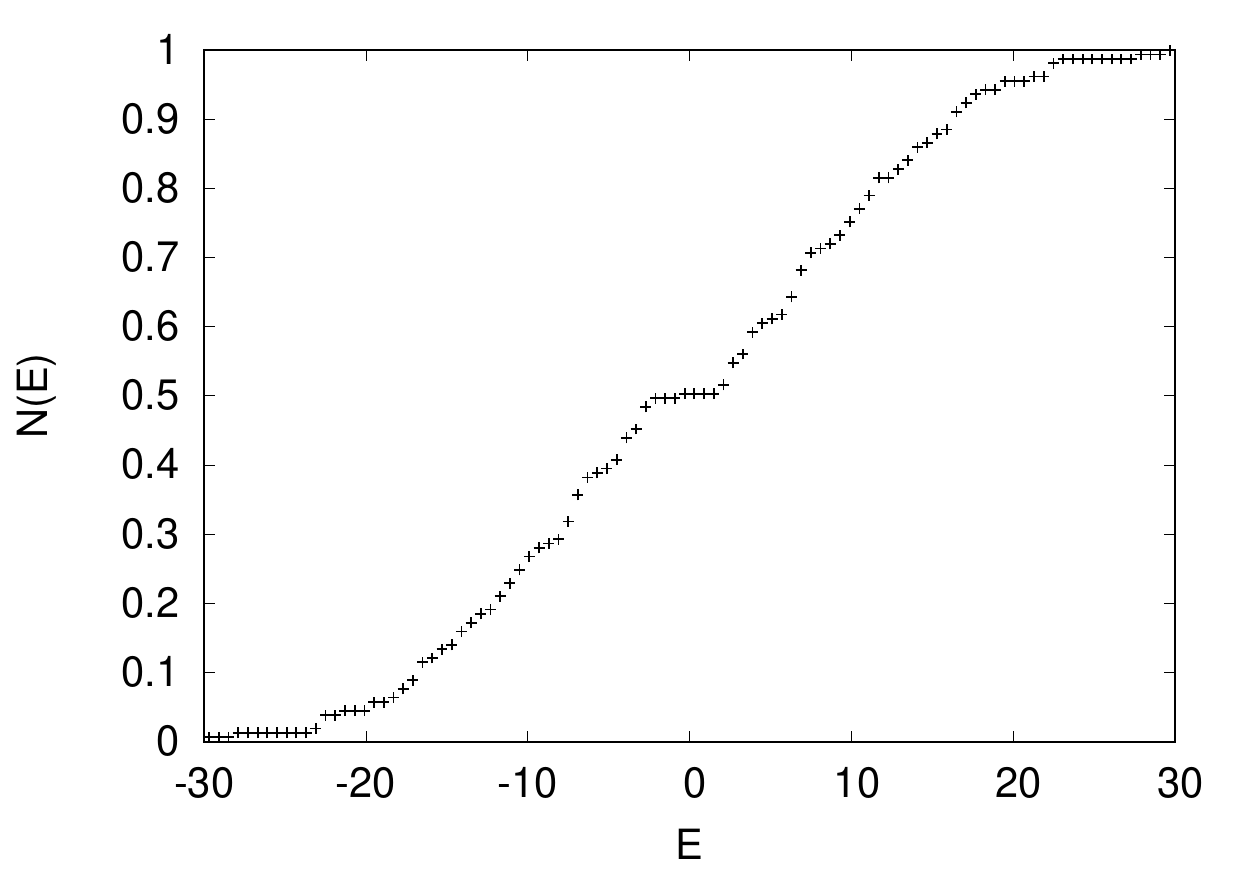}
\caption{The integrated d.o.s plot after degeneracies have been removed. This is the data that we use for doing the unfolding.}
\label{idosnondegplot}
\end{figure}

\begin{figure}[h]
\centering
\includegraphics[width=0.5\textwidth]{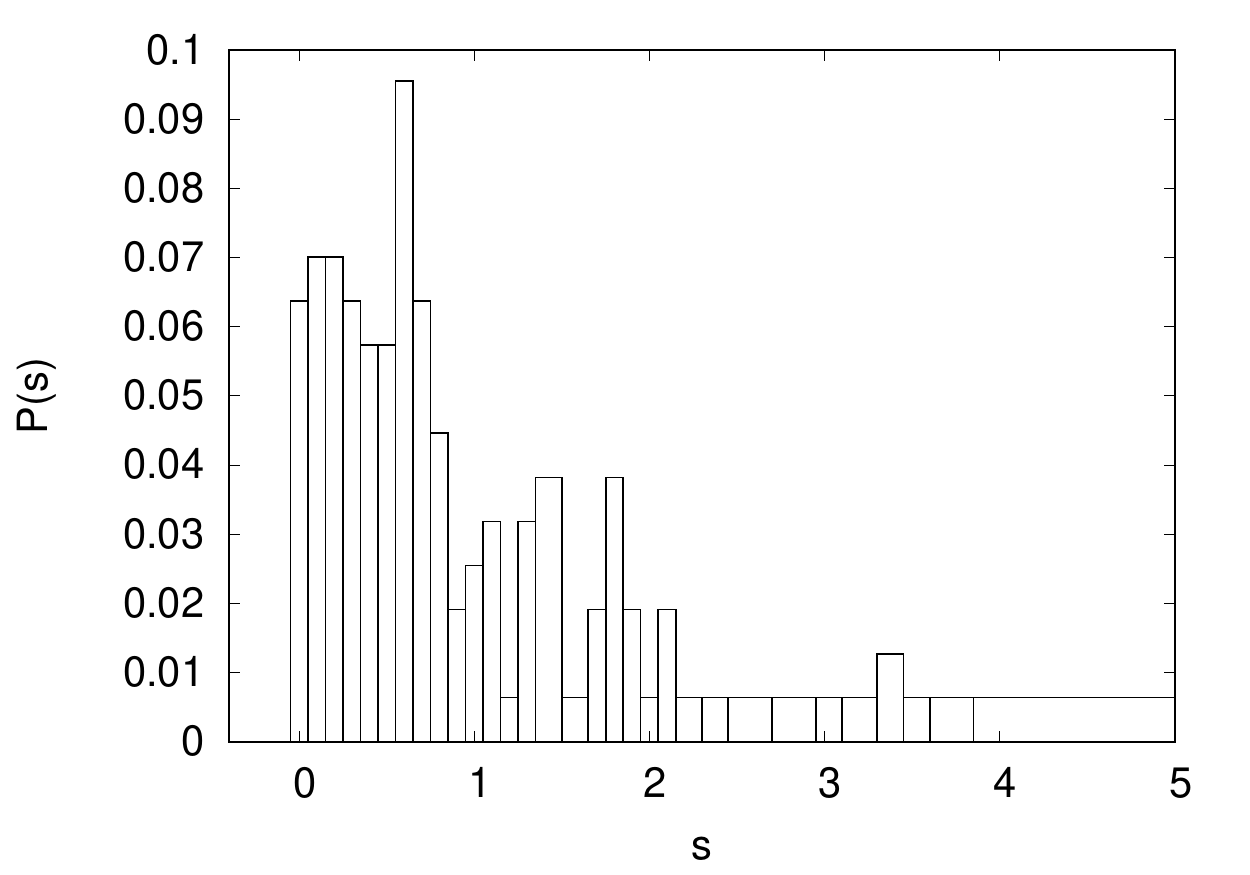}
\caption{Unfolded level spacing distribution showing level repulsion near $s\rightarrow 0$. The level repulsion is evident, but we emphasize that after the degeneracies are removed, the eigenvalues available are not many.}
\label{unfold}
\end{figure}



\section{Discrete Symmetries and the Choice of Ensemble}

From a glance at the spectrum, it becomes clear that the eigenvalues are exactly symmetrical around zero. Such a spectrum is said to exhibit {\em spectral mirror symmetry} \cite{Shukla}. In this section we will understand this symmetry in the spectrum in terms of an underlying discrete symmetry of the system. This will enable us to also identify the ensemble that is likely to control the random matrix-like behavior of the $D=3, n=2$ Gurau-Witten theory.

The basic observation here is simple. We note that flipping the sign of any one of the $\psi_a$'s in the theory changes the sign of the Hamiltonian: there is a unitary \cite{Witten} operator under which the Hamiltonian is odd. Following the conventions of \cite{Shukla}, we will call this the $S$ operator. The statement then is that
\bea
S H S^\dagger= -H
\eea
What is this operator explicitly? It is straightforward to see this in the gamma matrix language. Flipping $\psi_0$ corresponds in this language to flipping the signs of all the $\gamma_i$'s in the range $i=1, ..8$ while retaining the signs of all the rest\footnote{Flipping the signs of any of the other $\psi_a$'s can be understood as a (signed) permutation of the $\psi_a$'s together with the $S$ operation, and the former is a symmetry of the theory, so these do not give rise to essentially new $S$ operators.}. This means that $S$ is defined by
\bea
S= \gamma_1 \gamma_2 ... \gamma_8
\eea
so that 
\bea
S \gamma_i S = -\gamma_i \  {\rm for} \ i=1,...,8 \\
S \gamma_i S = +\gamma_i \ {\rm for} \  i=9,...,32
\eea
Note also that in the really real representation that we are working with, the gamma's are real and symmetric and so the Clifford algebra guarantees that $S^2=S S^{\dagger}=S S^T=1$. So what we are left with is a unitary operator $S$ that anti-commutes with the Hamiltonian, and squares to 1.

Furthermore, it was noted in \cite{Ludwig, Fu, crowdsourced} that the theory has a symmetry $P$ that has been called a particle-hole symmetry\footnote{It is perhaps more usefully called a $T$ operator. We will adopt this terminology. It contains an anti-linear piece and is related to Kramer's degeneracy, see page 10 of \cite{Fu}.}. The same construction goes through in our case as well. For SYK with $N=0$ mod 8, as well as in our case, it is straightforward to check that it squares to 1. 

Together then, we have two discrete symmetries. An $S$ that squares to 1, and a $T$ that squares to 1. It turns out that these two symmetries are the defining features of the symmetry class BDI in the Altland-Zirnbauer 10-fold classification. It is also referred to as the chiral Gaussian Orthogonal Ensemble. This observation is a strong suggestion that unlike in the SYK cases, the random matrix ensembles corresponding to the Holographic Tensor Models need not be the Wigner-Dyson ensembles.   
\begin{figure}[h]
\centering
\includegraphics[width=0.5\textwidth]{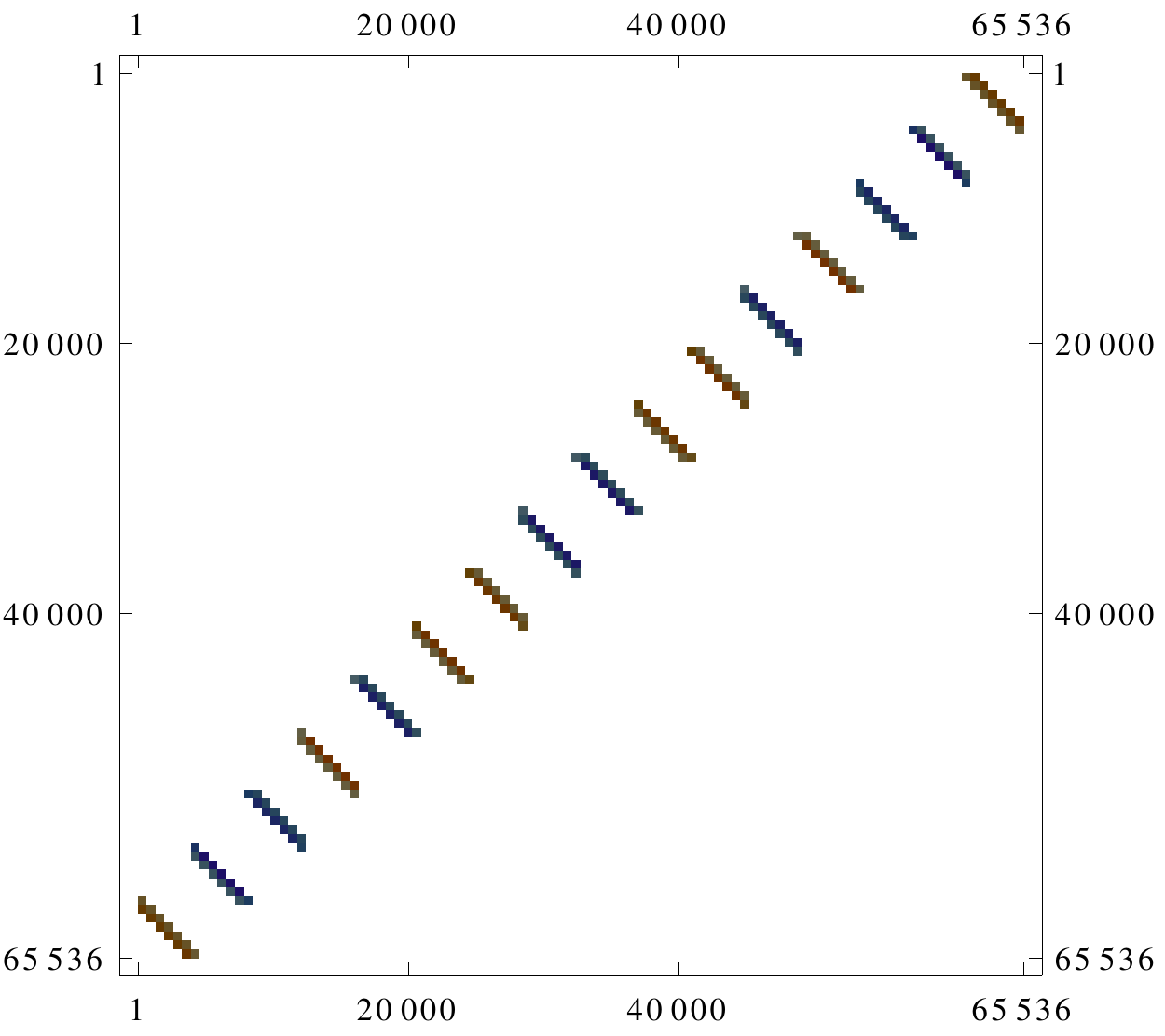}
\caption{The MatrixPlot structure of S.}
\label{Splot}
\end{figure}

We conclude this section with one brief comment. Note that Figure \ref{hamplot} is very suggestive of a Bogolubov-de Gennes (BdG) structure for the Hamitonian. This structure refers to Hamiltonians of the form
\bea
H = \begin{pmatrix}
\; A & B \;\\
\; B^{\dagger} & -A^{T}\; 
\end{pmatrix}
\eea
which are common in mesoscopic physics. One can in fact check explicitly that our Hamiltonian actually satisfies $A=A^T$.
Also since the Hamiltonian is real symmetric, we also have $B^\dagger=B^T$. But our Hamiltonian does {\em not} satisfy $B=\pm B^T$ which would have taken it to one of the other symmetry classes instead of BDI. Operationally this is because the $S$ operator in our case is {\em not} of the form 
\bea
\begin{pmatrix}
\; 0 & 1 \;\\
\; \pm 1 & 0\; 
\end{pmatrix},
\eea
see for example \cite{Shukla}. Explicit evaluations shows that its structure is as in Figure \ref{Splot} in the gamma matrix representation that we are working with.

\section{Comments}

Clearly, we have only considered the most basic features of a specific holographic tensor model. The results we find are a strong suggestion that there is a lot to be understood here. We only make some brief comments of immediate relevance.

It will be very interesting to understand the detailed level spacing distribution and other "random matrix-like" quantities of  HTMs with larger $N$: in our $N=32$ case we do not have too much statistics once the degeneracies are removed because the total number of eigenvalues of the Hamiltonian is merely 65536. The next simplest Gurau-Witten model however is at $D=3, n=3$ and $D=5, n=2$ which corresponds to $N=(D+1) n^D=108$ and $192$ which is computationally inaccessible via brute force\footnote{We are informed by J. Sonner that one can avoid dealing with explicit matrix assignments for gamma matrices, by treating operations involving them as logical operations on their matrix elements. This will reduce some of the demands on computing.}. Another possibility is to consider the model considered in \cite{klebanov}\footnote{We have completed a similar investigation there as well [19], and the results are quite parallel. In the $n=3, D=3$ case ($N=27$), there is spectral mirror symmetry and the plots of the various quantities are qualitatively similar. (Except for an overall 16-fold degeneracy that can be understood from symmetries.). The $n=2, D=3$ case ($N=8$) turns out to be too small to exhibit chaos. We have not been able to diagonalize the $n=2, D=5$ case ($N=32$) because the matrix is too dense for our rather simple-minded endeavors in computing.}, where the model is uncolored and therefore one gets a reduction in degrees of freedom by a factor of $D+1$. The $N$-dependence of the various features would be interesting to understand.

One thing we have not emphasized in this paper is the existence of the $\sim O(N)^{D(D+1)/2}$ symmetry in the Gurau-Witten theory, which should appropriately be thought of as gauged for holographic purposes. We have limited our discussion to a direct comparison with the SYK model where this symmetry is absent. See discussion in \cite{klebanov} for comments on this. 

We have also done some partial investigations of the thermodynamics of this model, but a thorough discussion will be presented elsewhere. It might also be interesting to see if the Gurau-Witten model is exactly solvable at large $D$.

\section*{Acknowledgement}

CK thanks Prithvi Narayan, Julian Sonner and Junggi Yoon for highly educational lectures on the SYK model and related matters. PNBS thanks Adhip Agarwala for useful discussions. We also thank Pallab Basu for collaboration in the early stages of this project. The use of the computational resources of IISc and especially the cluster at ICTS is acknowledged. SS gratefully acknowledges funding from Indo-Israeli joint research grant.  We also thank Pallab Basu, K. V. Pavan Kumar and R. Loganayagam for discussions. CK is grateful to Taamarakkutti for bringing order to his daily hustle during this project.

\end{document}